\documentclass[12pt]{article}

\begin{document}

\begin{center}

{\Large {{\bf{Scalar field on AdS: \\
quantum one loop "in one line"}} \\
Report at the Ginzburg Centennial Conference \\
Lebedev Institute \\
Moscow, 29 May - 03 June 2017

\vspace{1cm}

Boris L. Altshuler\footnote{E-mail addresses: baltshuler@yandex.ru $\,\,\,  \& \,\,\,$  altshul@lpi.ru}

\vspace{0,5cm}

{\it Theoretical Physics Department, \\
P.N. Lebedev Physical Institute, \\
53 Leninsky Prospect, Moscow, 119991, Russia}}}

\vspace{1cm}

\end{center}

{\bf Abstract:} It is shown that quantum one-loop potentials of the bulk fields in the Randall and Sundrum (RS) model may be immediately expressed in integrals with use of Barvinsky-Nesterov or equivalently Gelfand-Yaglom methods of calculation of quantum determinants. Simple expression is obtained for the UV-finite difference of one-loop quantum energies for two arbitrary values of  parameter of the double-trace asymptotic boundary condition in one-brane and two-brane RS-models. Compact formula for Schwinger-DeWitt expansion gives values of induced Planck mass and induced gauge coupling constant in transverse space. S-DW expansion is plagued by IR-divergences in one-brane RS-model and makes sense in two-brane model.

\newpage

\tableofcontents

\newpage

\section{Preliminaries and main results}

\subsection{Preliminaries}

\quad This report continues the research of the author's papers \cite{Alt2017} (2017) and \cite{Alt2015} (2015). We again consider $(d+1)$-dimensional AdS space of the Randall and Sundrum model in Euclidean Poincare coordinates:

\begin{equation}
\label{1}
ds^{2}= \frac{dz^{2}+\tilde{g}_{\mu\nu}(x)dx^{\mu}dx^{\nu}}{(kz)^{2}},
\end{equation}
where $\epsilon < z < L$ ($z=\epsilon, \, L$ are positions of UV and IR branes), $\mu, \nu = 0,1,...(d-1)$, $\tilde{g}_{\mu\nu}(x)=\delta_{\mu\nu}$ in all sections but one dedicated to Schwinger-DeWitt expansion, $k$ is AdS scale. Bulk scalar field $\Phi=\Phi(\vec{p},z)$ "living" on this space satisfies equation ($\vec{p}$ is momentum in transverse $d$-space):

\begin{equation}
\label{2}
\hat{D}(p) \Phi = \left[-z^{2}\frac{\partial^{2}}{\partial z^{2}}+(d-1)z\frac{\partial}{\partial z}+\left(\nu^{2}-\frac{d^{2}}{4}\right)+z^{2} p^{2}\right]\,\Phi = 0.
\end{equation}

In fact all formulas of the Report may be directly applied for any value of non-integer $\nu$ (the case of integer $\nu$ will be discussed in Conclusion).

\subsection{Hypothesis on higher spins.}

Solution of Eq. of motion of field of any spin on $AdS$ space has a form {\footnote{This follows from the most general group-theoretical considerations. I am grateful to Ruslan Metsaev for clarifying this point.}}:

\begin{equation}
\label{3}
\phi_{\mu_{1}...\mu_{s}} \sim z^{\gamma}I_{\pm\nu}(pz).
\end{equation}
The value of $\gamma$ is irrelevant for us; for Bose field of spin $s > 0$ $\nu$ is equal to

\begin{equation}
\label{4}
\nu = \sqrt{\frac{m^{2}}{k^{2}} + \left(s + \frac{d-4}{2}\right)^{2}}.
\end{equation}
For Fermi fields of spin $s$ there are also quite simple expressions for $\nu$ \cite{Metsaev}.

Thus we may suppose that substitution of the proper values of $\nu$ in the formulas of the Report will give quantum one-loop potentials and Schwinger-DeWitt expansion for the fields of any spin $s$ on the background of the one-brane or two-branes Randall-Sundrum model.

This hypothesis needs to be verified. (See also Sec. 1.4 below).

\subsection{Asymptotic b.c.}

General solution of Eq. (\ref{2}):

\begin{equation}
\label{5}
\Phi(z) = z^{d/2}[C_{1}I_{\nu}(pz) + C_{2}I_{-\nu}(pz)] \to \alpha(p)\,z^{\frac{d}{2}+\nu} + \beta(p)\,z^{\frac{d}{2} - \nu}
\end{equation}
at $z \to 0$. In the framework of the $AdS/CFT$ correspondence on the gravity side $\alpha(p)$ plays the role of the source of single-trace operator $\hat{O}$ of the boundary CFT whereas $\beta(p)$ is its quantum average.

Witten showed in 2001 \cite{Witten} that deformation of the Action of the boundary CFT by the multi-trace term $W(\hat{O})$ corresponds on the gravity side to the asymptotic b.c. $\alpha = \partial W(\beta)/ \partial \beta$. Double-trace deformation $W = (1/2)f \hat{O}^{2}$ gives $\alpha = f \beta$. Corresponding Green function looks as \cite{Mitra}:

\begin{eqnarray}
\label{6}
G_{f}(p;z,z')= - k^{d-1}(zz')^{d/2}\, \cdot   \qquad \qquad  \nonumber
\\ \nonumber
\\
\cdot \frac{\{[I_{-\nu}(pz)+\bar{f}(p)I_{\nu}(pz)]\,K_{\nu}(pz')\,\theta(z'-z)+ (z \leftrightarrow z')\}}{1+\bar{f}(p)}
\\  \nonumber
\\   \nonumber
\bar{f}(p) = f\,\left(\frac{2}{p}\right)^{2\nu}\,\frac{\Gamma (1+\nu)}{\Gamma (1-\nu)}. \qquad  \qquad  \qquad  \nonumber
\end{eqnarray}
here $L = \infty$; $\bar{f}(p)$ is obtained from comparison of asymptotic of $u_{f}(z) = z^{d/2}[I_{-\nu}(pz)+\bar{f}(p)I_{\nu}(pz)]$ with asymptotic (\ref{5}) with account of b.c. $\alpha = f\beta$.

Difference of Green functions for two values of $f$: $G_{f_{2}} - G_{f_{1}} \sim (f_{2}- f_{1})K_{\nu}(pz)K_{\nu}(pz')$ is UV-finite at coinciding arguments \cite{Mitra}:

\begin{equation}
\label{7}
\int [G_{f_{2}}(p;z,z) - G_{f_{1}}(p;z,z)]\,d^{d}p  \, < \, \infty.
\end{equation}

Also Gubser \& Mitra \cite{Mitra} (see also Diaz \& Dorn (2007) \cite{Diaz} and Hartman \& Rastelli (2008) \cite{Hartman}) calculated the UV-finite difference $V^{(d)}_{+}- V^{(d)}_{-}$ of quantum potentials for regular ($f_{2} = \infty$) and irregular ($f_{1} = 0$) Green functions. However, as it was pointed out in \cite{Mitra}, it is hard to calculate this difference for general values of $f$. {\bf{This problem is resolved in the present report}}.

\subsection{Doubtfulness of expressions for $a$-anomaly obtained for zero spin by Mitra \& Gubser (2002) \cite{Mitra} (and also by Diaz \& Dorn (2007) \cite{Diaz} and Hartman \& Rastelli (2008) \cite{Hartman}), and for arbitrary spin by Giombi, Klebanov, Pufu, Safdi, \& Tarnopolsky (2013) \cite{Giombi}}

In general for difference of quantum potentials corresponding to ratio of determinants for two b.c. we have ($[F]_{2-1} \equiv F_{2} - F_{1}$):

\begin{eqnarray}
\label{8}
[V^{(d)}]_{2-1} = \frac{1}{2}\int\,\frac{d^{d}p}{(2\pi)^{d}}\, \ln \left[\frac{{\rm Det}_{2}\hat{D}(p)}{{\rm Det}_{1}\hat{D}(p)}\right] \stackrel{?}{=} \frac{1}{2} \int^{m^{2}} d\tilde{m}^{2} {\rm{Tr}}\left[G(\tilde{m}^{2})\,\frac{\partial{{\hat{D}}}}{\partial{\tilde{m}^{2}}}\right]_{2-1} = \nonumber
\\
\\
= \frac{1}{2} \int^{m^{2}} d\tilde{m}^{2}\,\int_{\epsilon}^{L}\,dz\,\int\,\frac{d^{d}p}{(2\pi)^{d}}\,\left[G(p;z;z;\tilde{m}^{2})\,\frac{\partial{{\hat{D}}}}{\partial{\tilde{m}^{2}}}\right]_{2-1}. \qquad  \qquad  \qquad  \nonumber
\end{eqnarray}

B-N or equivalently G-Y methods used in the report immediately give simple formulas for ratio of determinants in (\ref{8}), that is they {\bf{permit to "jump over" two integrations}} (over $m^{2}$ and over $z$) in (\ref{8}).

Now I comment the question mark over equality in (\ref{8}).

There is no objections to the nice formula for difference of "regular" and "irregular" Green functions at coinciding arguments obtained in \cite{Mitra} - \cite{Hartman} for spin zero and in \cite{Giombi} for any spin; for even spin it is given by the integrand of formula (9.2) of \cite{Giombi} which for particular case $d=4$ is equal to:

\begin{equation}
\label{9}
G^{(4)}_{+}(x,z;x,z) - G^{(4)}_{-}(x,z;x,z) \sim (s+1)^{2}\,\nu\,[\nu^{2} - (s+1)^{2}],
\end{equation}
here $\nu = \Delta - d/2 = \Delta - 2$. For $s=0$  this reproduces result of papers \cite{Mitra} - \cite{Hartman} independently obtained by Saharian (2005) \cite{Saharian}.

Doubtful is the calculation of the $a$-anomaly coefficient (proportional to the difference of quantum energies) with integration of (\ref{9}) over mass squared $\int\,dm^{2} = \int\,2\nu\,d\nu$ \cite{Mitra} - \cite{Giombi}. In particular for $d = 4$ this integration gives Eq. (9.3) of \cite{Giombi} (where we again changed $\Delta - 2 \to \nu)$:

\begin{equation}
\label{10}
V^{(4)}_{+} - V^{(4)}_{-}  \sim a^{(s)}_{UV} - a^{(s)}_{IR} \sim (s + 1)^{2}\,\nu^{3}[5\,(1 + s)^{2} - 3\,\nu^{2}].
\end{equation}

For $s = 0$ this result coincides with one of \cite{Mitra} - \cite{Hartman}, but drastically differs from the results (also coinciding) of \cite{Alt2015} and \cite{Saharian} obtained in different ways (Saharian \cite{Saharian} obtained expression for LHS of (\ref{10}) by the direct calculation of the energy-momentum tensor with use of Green functions (\ref{9})).

To our mind the pitfall is in the use of the familiar trick:

\begin{equation}
\label{11}
\ln{\rm{Det}}\hat{D} = \int\,dm^{2}\,\frac{\partial \ln{\rm{Det}}\hat{D}}{\partial m^{2}} \stackrel{?}{=} \int\,dm^{2}\,{\rm{Tr}}\,\left[G\,\frac{\partial{\hat{D}}}{\partial m^{2}}\right],
\end{equation}
where in \cite{Mitra} - \cite{Giombi} $\partial\hat{D} / \partial{m^{2}} = 1$ everywhere. This however is valid only in situations when boundary conditions (which determine the spectra and which are hidden in the LHS and RHS of (\ref{11})) do not depend on $m^{2}$.

This is actually just the case for ordinary b.c. with fixed Robin coefficients. But we shall see below that for asymptotic b.c., including the "regular" (+) and "irregular" (-) cases, effective Robin coefficients depend on $\nu$ that is on $m^{2}$. Thus when asymptotic b.c. are imposed RHS of (\ref{11}) (and of (\ref{8})) must be modified with local boundary terms of $\partial\hat{D} / \partial{m^{2}}$ which are not taken into account in \cite{Mitra} - \cite{Giombi}. We shall clear up it below in subsection 3.2.

Thus we shall go another way which immediately gives ratio of determinants in the LHS of (\ref{8}) without calculation of difference of Green functions.

Looking forward we present now {\bf{the expression for $a$-anomaly in the one-brane RS-model obtained below in the Report}} ($a_{UV} - a_{IR} \sim V^{(d)}_{+} - V^{(d)}_{-}$):

\begin{equation}
\label{12}
{\tilde{V}}^{(d)}(\nu) = V^{(d)}_{+} - V^{(d)}_{-} = \frac{2\sin(\pi\nu)\,\Omega_{d-1}}{(2\pi)^{d+1}d\,\epsilon^{d}} \int_{0}^{\infty}\frac{y^{d-1}dy}{I_{\nu}(y)\,I_{-\nu}(y)}.
\end{equation}

For higher spin fields the factor $g(s)$ (number of degrees of freedom - see e.g. (6.15) in \cite{Giombi}) must be included in the RHS of (\ref{12}). For massless high spin fields $\nu$ is integer or half-integer. The case of integer $\nu$ must be considered separately - see discussion in the Conclusion. It would be nice to have the analytical formula for ${\tilde{V}}^{(d)}(s)$ (\ref{12}) for half integer $\nu = s = n + 1/2$ ($n = 0, 1, 2...$). {\bf{I did not manage to find such a formula for definite integral in (\ref{12}).}}

\subsection{Main results of the Report}

\quad (1) simple expressions for ratios of quantum ${\rm{Det}}\hat{D}$ for different Robin or asymptotic b.c., and for corresponding one-loop potentials are obtained;

(2) Schwinger-DeWitt expansion for metric $\tilde{g}_{\mu\nu}(x)$ of the transverse space in (\ref{1}) is put down, induced Planck mass and gauge coupling are calculated.

\section{Equivalence of G-Y and B-N methods}

\subsection{Gelfand-Yaglom (G-Y) method (1960) \cite{GY} - \cite{Kirsten2}}

G-Y method says that in the one-dimensional problem $\hat{D}\Phi_{n}(t) = \lambda_{n}\Phi_{n}(t)$ ($a<t<b$) determinant ${\rm{Det}}\hat{D}$ for b.c. $\dot{\Phi}(a) + r_{a}\Phi(a) = 0$, $\dot{\Phi}(b) + r_{b}\Phi(b)=0$ ($\dot{\Phi} \equiv \partial / \partial{t}$), may be expressed through solution $v(t)$ of homogeneous equation $\hat{D}v(t) = 0$ which obeys b.c. at one boundary, say at $t=b$. Then ${\rm Det}\hat{D}$ is given by the combination of other b.c. built with this $v(t)$:

\begin{eqnarray}
\label{13}
{\rm Det}\hat{D} \sim \dot{v}(a) + r_{a}v(a) \qquad  {\hat{D}}v(t)=0 \qquad  \dot{v}(b) + r_{b}v(b)=0, \nonumber
\\
\\
{\rm Det}\hat{D} \sim \dot{u}(b) + r_{b}u(b) \quad  {\hat{D}}u(t)=0 \qquad  \dot{u}(a) + r_{a}u(a)=0. \quad \nonumber
\end{eqnarray}

Let us outline in short the proof of (\ref{13}) which is double-step:

(1) for solution $\phi(z|\lambda)$ of Eq. $\hat{D}\phi = \lambda\,\phi$, which obeys b.c. at one end, say at $t=b$: $[\dot{\phi}(b|\lambda) + r_{b}\phi(b|\lambda)]=0$, function $B(\lambda) \equiv [\dot{\phi}(a|\lambda) + r_{a}\phi(a|\lambda)]$ have zeroes at $\lambda = \lambda_{n}$: $B(\lambda_{n}) = 0$;

(2) since logarithmic derivative of $B(\lambda)$ has poles in complex $\lambda$-plane exactly at $\lambda = \lambda_{n}$ it is possible to express $\zeta$-function ($\zeta (s) = \sum{\lambda_{n}}^{-s}$) with contour integral over this logarithmic derivative and finally, after a number of rather conventional steps, to get the looked for G-Y formula $e^{-\zeta'(0)}={\rm{Det}}\hat{D} \sim B(\lambda=0) = [\dot{\phi}(a|0) + r_{a}\phi(a|0)]$ which is exactly the first line in (\ref{13}) (since $\phi(z|0)$ is a homogeneous solution $v(z)$ obeying b.c. at $t=b$).

{\bf{Note}}: Wronskian of two solutions $u(t)$, $v(t)$ introduced in (\ref{13}) $W(u,v) = \dot{u}\,v - \dot{v}\,u$ is proportional to their corresponding G-Y combinations:

\begin{equation}
\label{14}
W(u,v) = -u(a)[\dot{v}(a)+r_{a}v(a)] = v(b)[\dot{u}(b) + r_{b}u(b)],
\end{equation}
($\dot{u}(a) = -r_{a}u(a)$; $\dot{v}(b) = - r_{b}v(b)$ were used in $W(u,v)$ in (\ref{14})). This elementary observation will permit us to demonstrate the equivalence of G-Y and B-N methods when ratios of determinants for different b.c. are calculated.

\subsection{Warm-up. Casimir potential in one line: a few inspiring examples.}

Let us show how G-Y method immediately gives familiar results obtained conventionally in a rather lengthy way. Examples considered below refer to flat $(d+1)$-dimensional space and to elementary differential operator

\begin{equation}
\label{15}
\hat{D}_{0}(p) = - \frac{\partial^{2}}{\partial t^{2}} + p^{2}.
\end{equation}

1. Classical Dirichlet-Dirichlet problem ($0 < t < L$): $\phi(0)=0$, $\phi(L)=0$. $v=C\cdot \sinh(pt)$ is a solution of Eq. $\hat{D}_{0}\phi = 0$ satisfying b.c. at $t=0$. Then according to G-Y method ${\rm Det}_{D-D}\hat{D}_{0} \sim \sinh(pL)$. This yields expression for quantum potential in $d$ dimensions: $V_{D-D}^{(d)} = $

\begin{equation}
\label{16}
= \frac{1}{2}\int \frac{d^{d}p}{(2\pi)^{d}}\,\ln\,[\sinh(pL)] = A + BL - \frac{1}{L^{d}}\frac{\Omega_{d-1}}{(2\pi)^{d}\,2^{d+1}\,d}\,\int_{0}^{\infty}\frac{y^{d}dy}{e^{y}-1},
\end{equation}
where $A$, $B$ are irrelevant divergent constants. Last term is the Casimir potential which is zero at $L \to \infty$ and which is UV-finite. It is easy to check that this expression gives its well known values in (1+1) and in (3+1) dimensions: $V_{Cas\, D-D}^{(1)}L= - \pi/24$, $V_{Cas \, D-D}^{(3)}L^{3}= - \pi^{2}/1440$ (for electromagnetic field this result must be multiplied by 2 - number of polarizations of e-m field).

2. $M^{d}\times S^{1}$ where $S^{1}$ is a circle of length $L=2\pi\rho$. In this case spectra of periodic (untwisted) or antiperiodic (twisted) modes are found from

\begin{equation}
\label{17}
\cos (\sqrt{\lambda_{n}- p^{2}}\,L) = \pm 1.
\end{equation}

Then  in untwisted case for example, ${\rm Det}_{\it{untw}}\hat{D}(p) \sim (\cosh pL - 1)$. Thus:

\begin{equation}
\label{18}
V^{(d)}_{\it{untw}} = \frac{1}{2}\int \frac{d^{d}p}{(2\pi)^{d}}\,\ln\,[\cosh(pL) - 1] = ... - \frac{1}{L^{d}}\frac{\Omega_{d-1}}{(2\pi)^{d}\,\,d}\,\int_{0}^{\infty}\frac{y^{d}dy}{e^{y}-1},
\end{equation}
where {\it{dots}} symbolise irrelevant divergent terms whereas the last expression gives Casimir vacuum energy $V^{(d)}_{Cas,\it{untw}}$ which is UV-finite and is equal to zero at $L \to \infty$. It gives in particular well known result for torus in 5 dimensions, i.e. for $d=4$: $V^{(4)}_{Cas\, {\it{untw}}} \cdot \rho^{4} = - 3\zeta(5) / (2\pi)^{6}$, obtained by Candelas \& Weinberg (1984) \cite{Candelas} with rather complex calculations. It is easy to get in the same way well known values of Casimir potential for twisted modes.

3. In the same way the massless version of the final formula (22) of paper \cite{Elizade} (Elizalde, Odintsov \& Saharian - 2009) for Casimir potential for most general Robin b.c. on both boundaries in flat space is obtained "in one line".

The generalization of these examples to massive scalar field is obvious.

{\bf{Now we come to the ratios of quantum determinants of one and the same differential operator and different b.c.}}

\subsection{Barvinsky-Nesterov (B-N) method (2005) \cite{Barv2005}-\cite{Barv2014}}

B-N method roots in the fact that ${\rm{Det}}\hat{D}$ is given by a product of functional integrals over bulk field $\bar{\Phi}$ and over boundary field $\Phi_{b}$:

\begin{equation}
\label{19}
[{\rm{Det}}\hat{D}]^{-1/2} = \int\,D\Phi\,e^{\Phi\,\hat{D}\,\Phi}= \int_{b}\,D\Phi_{b}\,e^{\Phi_{b}\,{\hat{F}}_{b}\,\Phi_{b}} \cdot \int_{\Phi = \Phi_{b}}D\bar{\Phi}\,e^{\bar{\Phi}\,{\hat{D}}_{bulk}\,\bar{\Phi}},
\end{equation}
where "Robin mass terms" are included in the boundary action. Thus in ratio of determinants for two different b.c. bulk functional integrals reduce, and this ratio is given by the ratio of determinants of B-N boundary operators:

\begin{equation}
\label{20}
\frac{{\rm{Det}}_{2}\hat{D}}{{\rm{Det}}_{1}\hat{D}} = \frac{{\rm{det}}\hat{F}_{2b}}{{\rm{det}}\hat{F}_{1b}}.
\end{equation}

\begin{equation}
\label{21}
\hat{F}_{b} = - \hat{G}_{Dir\,,tt'}(t,t'|x,x')_{||} + \hat{r}(x,x') = [\hat{G}_{r_{a},r_{b}}(t,t'|x,x')_{||}]^{-1},
\end{equation}
$\hat{F}_{b}$, $\hat{G}_{Dir}$, $\hat{r}$, $\hat{G}_{r_{a},r_{b}}$ are (2 x 2) matrixes which elements are operators in $x^{\mu}$ space; symbol $||$ means that $t,t'$ are taken at the boundaries $t\,t' = a \,b$; $\hat{r} = diag(r_{b}; - r_{a})$. $\hat{G}_{r_{a},r_{b}}$ is Green function obeying Robin b.c. at $t\,t' = a \,b$.

Last equality in (\ref{21}) is crucial in B-N approach, its general proof is given in \cite{Barv2005}. However this equality may be demonstrated easily in one-dimensional problem parametrized by transverse momentum $\vec{p}$ when elements of (2 x 2) matrixes in (\ref{21}) are numbers.

In the one-dimensional problem Green function $G_{r_{a},r_{b}}(t,t';p)$ is built in a standard way with functions $u(t)$, $v(t)$ (\ref{13}):

\begin{equation}
\label{22}
G_{r_{a},r_{b}}(t,t';p) = \frac{u(t)\,v(t')\,\theta(t'-t)+ (t \leftrightarrow t')}{\dot{u}(t)\,v(t) - \dot{v}(t)\,u(t)}
\end{equation}

Determinant of corresponding (2 x 2) matrix ${\hat{G}}_{r_{a}, r_{b}}(t,t';p)_{||}$:

\begin{equation}
\label{23}
{\rm{det^{(2 x 2)}}}{\hat{G}}_{r_{a},r_{b}}(t,t';p)_{||} = - \frac{u(a)\,v(b)}{W(u,v)}\cdot v_{D}(a) = [{\rm{det}}\hat{F}_{b}]^{-1},
\end{equation}
where $W(u,v)$ is Wronskian, $v_{D}(t)$ is a solution of Eq. $\hat{D}v(t) = 0$, satisfying b.c.: $v_{D}(b) = 0$, $v'_{D}(b) = 1$. Thus summing up Eqs. (\ref{20}), (\ref{21}), (\ref{23}) the demanded ratio of one-loop quantum determinants is finally obtained:

\begin{eqnarray}
\label{24}
\frac{{\rm{Det}}_{2}\hat{D}}{{\rm{Det}}_{1}\hat{D}} = \frac{{\rm{det}}\hat{F}_{2b}}{{\rm{det}}\hat{F}_{1b}} = \frac{{\rm{det}}\hat{G}_{r_{1a},r_{1b}}(t,t';p)_{||}}{{\rm{det}}\hat{G}_{r_{2a},r_{2b}}(t,t';p)_{||}} = \qquad \qquad   \qquad  \nonumber
\\
\\
= \frac{u_{r_{1a}}(a)\,v_{r_{1b}}(b)}{W(u_{r_{1a}},v_{r_{1b}})} \cdot \frac{W(u_{r_{2a}},v_{r_{2b}})}{u_{r_{2a}}(a)\,v_{r_{2b}}(b)} = \frac{[\dot{v}_{r_{2b}}(a)+r_{2a}v_{r_{2b}}(a)]\,v_{r_{1b}}(b)}{[\dot{v}_{r_{1b}}(a)+r_{1a}v_{r_{1b}}(a)]\,v_{r_{2b}}(b)}.   \nonumber
\end{eqnarray}
Here last equality (obtained from (\ref{14})) demonstrates the equivalence of B-N and G-Y methods.

{\bf{Expression (\ref{24}) is the main working tool of the report.}}

\section{Ratio of determinants in RS-model}

\subsection{Two asymptotic b.c.}

From now on we consider one and the same Robin b.c. at $z = L < \infty$, and different asymptotic b.c. $\alpha = f_{1,2} \beta$ at $z\to 0$. Differentiation $\partial / \partial t$ in previous sections now in Poincare coordinates (\ref{1}) comes to $z\,\partial /\partial z$.

Corresponding Green function is built of Gubser \& Mitra function $u_{f}$ obeying asymptotic b.c. $\alpha = f\,\beta$ (cf. (\ref{6})):

\begin{equation}
\label{25}
u_{f}(z) = z^{d/2}[I_{-\nu}(pz)+\bar{f}(p)\,I_{\nu}(pz)]; \quad \bar{f}(p) = f\,\left(\frac{2}{p}\right)^{2\nu}\,\frac{\Gamma (1+\nu)}{\Gamma (1-\nu)},
\end{equation}
and of function $v_{r_{L}}(z)$ obeying Robin b.c. $[zv'(z) + r_{L}v(z)]_{z=L} = 0$:

\begin{eqnarray}
\label{26}
v_{r_{L}}(z)= z^{d/2}\,\left\{A_{r_{L}}[I_{\nu}(pL)]\,I_{-\nu}(pz) - A_{r_{L}}[I_{-\nu}(pL)]\,I_{\nu}(pz)\right\}, \qquad \nonumber
\\
\\ \nonumber
A_{r_{L}}[\psi(pz)] =\left(\frac{d}{2} + r_{L} \right)\,\psi (pz) + z\,\frac{\partial{\psi(pz)}}{\partial{z}}, \qquad  \qquad \qquad \qquad  \\ \nonumber
\end{eqnarray}

Wronskian $W(u_{f},v_{r_{L}}) = z\,u'_{f}v_{r_{L}} - u_{f}z\,v'_{r_{L}}$:

\begin{equation}
\label{27}
W(u_{f},v_{r_{L}}) = z^{d}\,\frac{2\sin \pi\nu}{\pi}\,\{A_{r_{L}}[I_{-\nu}(pL)] + \bar{f}(p)\,A_{r_{L}}[I_{\nu}(pL)]\}.
\end{equation}

Then according to the B-N prescription it is obtained:

\begin{eqnarray}
\label{28}
Q_{f_{2}f_{1}}(p) \equiv \frac{{\rm{Det}}_{f_{2}}\hat{D}}{{\rm{Det}}_{f_{1}}\hat{D}} = \frac{u_{f_{1}}(\epsilon)}{u_{f_{2}}(\epsilon)} \cdot \frac{W(u_{f_{2}},v_{r_{L}})}{W(u_{f_{1}},v_{r_{L}})} = \qquad \qquad  \qquad  \nonumber
\\
\\
= \frac{I_{-\nu}(p\,\epsilon) + \bar{f}_{1}(p)I_{\nu}(p\,\epsilon)}{I_{-\nu}(p\,\epsilon) + \bar{f}_{2}(p)I_{\nu}(p\,\epsilon)}\cdot \frac{A_{r_{L}}[I_{-\nu}(pL)] + \bar{f}_{2}(p)\,A_{r_{L}}[I_{\nu}(pL)]}{A_{r_{L}}[I_{-\nu}(pL)] + \bar{f}_{1}(p)\,A_{r_{L}}[I_{\nu}(pL)]}. \qquad \nonumber
\end{eqnarray}

{\bf{Two Notes:}}

1) $Q_{f_{2}f_{1}}(p) \to 1$ exponentially at $p \to \infty$. Hence $(V^{(d)}_{f_{2}} - V^{(d)}_{f_{1}})$ given by integral $\int\,d^{d}p\,\ln Q_{f_{2}f_{1}}(p)$ is UV-finite (surely this is not true  if $\epsilon = 0$).

2) RHS of (\ref{28}) depends on $\epsilon$ although there  is  no  hint  on $\epsilon$ neither in Witten's $\alpha = f\,\beta$, nor in Wronskian (\ref{27}) which zeroes (after substitution $p \to i\,\omega$) determine spectrum of operator $\hat{D}(p)$ (\ref{2}) corresponding to asymptotic b.c. $\alpha = f\beta$ at $z \to 0$ and to given Robin b.c. at $z=L$.

Then: Why $\epsilon$? Where does it come from?

The same question may be addressed to Mitra \& Gubser \cite{Mitra} who limited, just "by hand", integration over $z$ in ${\rm{Tr}}G$ (\ref{8}) with the lower limit $z = \epsilon$; that is why quantum energy in \cite{Mitra} depends on $\epsilon$ ($\sim \epsilon^{-d}$).

The same happens with the Barvinsky \& Nesterov {\it{definition}} of ${\rm{Det}}\hat{D} \sim {\rm{det^{(2 x 2)}}}\hat{F}_{b}$ where functional integral (\ref{19}) depends on positions of boundaries.

\subsection{Effective Robin coefficients}

As it was demonstrated above B-N method is equivalent to G-Y method. But what does it mean in case of asymptotic b.c.?

B-N ratio (\ref{28}) is equal to ratio of two G-Y combinations:

\begin{equation}
\label{29}
Q_{f_{2}f_{1}}(p) \equiv \frac{{\rm{Det}}_{f_{2}}\hat{D}}{{\rm{Det}}_{f_{1}}\hat{D}} = \frac{v'_{r_{L}}(\epsilon) + r_{\epsilon\,2}v_{r_{L}}(\epsilon)}{v'_{r_{L}}(\epsilon) + r_{\epsilon\,1}v_{r_{L}}(\epsilon)},
\end{equation}
where effective Robin coefficients $r_{\epsilon_{1,2}}$ are calculated from the identity:

\begin{eqnarray}
\label{30}
r_{\epsilon_{1,2}}(f_{1,2};p,\epsilon) \equiv - \frac{\epsilon \, u'_{f_{1,2}}(\epsilon)}{u_{f_{1,2}}(\epsilon)} = -\frac{d}{2} - \frac{\epsilon \,I'_{-\nu}(p\,\epsilon) + \bar{f}_{1,2}(p)\,\epsilon \,I'_{\nu}(p\,\epsilon)}{I_{-\nu}(p\,\epsilon) + \bar{f}_{1,2}(p)\,I_{\nu}(p\,\epsilon)} \approx  \qquad  \qquad \nonumber \\
\\  \nonumber
\stackrel{p\,\epsilon \ll 1}{\approx} - \frac{d}{2} + \nu \cdot \frac{1 - f_{1,2}\,\epsilon^{2\nu}}{1 + f_{1,2}\,\epsilon^{2\nu}} \qquad \left[\,\stackrel{p\,\epsilon \gg 1}{\to} - \frac{d}{2} - \frac{\epsilon\,I'_{\nu}(p\,\epsilon)}{I_{\nu}(p\,\epsilon)} \right],   \qquad  \qquad  \qquad \nonumber
\end{eqnarray}
where $u_{f}(z)$ is defined in (\ref{25}). We see that asymptotic of $r_{\epsilon}$ at $p\epsilon \gg 1$ ($[...]$ in (\ref{30})) does not depend on $f$!!! That is why RHS of (\ref{29}) $\to 1$ at $p \to \infty$ - like in (\ref{28}); hence corresponding quantum potential is UV-finite. This evidently is not the case of the conventionally fixed Robin coefficients $r_{\epsilon_{1,2}}$ in (\ref{29}).

Boundary conditions may be imposed dynamically \cite{Barv2005} with introduction of the Robin mass terms in the boundary action. Corresponding terms must be included in differential operator $\hat{D}(p)$ (\ref{2}). In our case this term in $\hat{D}(p)$ looks as $r_{\epsilon}\,\delta(z-\epsilon)$. And  since effective $r_{\epsilon}$ (\ref{30}) depends on $\nu$, that is on $m^{2}$, this term must be taken into account in $\partial{\hat{D}}/\partial\,m^{2}$ in (\ref{8}) or (\ref{11}). This was not taken into account in \cite{Mitra} - \cite{Giombi}; that is why their formulas for a-anomaly are doubtful.  It may be shown that RHS of (\ref{11}) (or of (\ref{8})) where local boundary terms of $\partial{\hat{D}}/\partial\,m^{2}$ are taken into account gives expressions for the one-loop quantum potentials obtained by B-N or G-Y methods.

\section{One loop for double-trace asymptotic $\alpha = f \, \beta$}

\subsection{Quantum potential for $f$ and possibility of the gap equation for $f$}

Let us put down difference of potentials corresponding to ratio $Q_{f_{2}f_{1}}(p)$ (\ref{28}) when $f_{1} = 0$ ("irregular" asymptotic b.c.) and $f_{2} \to f$ - arbitrary double-trace coefficient:

\begin{eqnarray}
\label{31}
\tilde{V}^{(d)}_{f,0}(\epsilon,L) \equiv V^{(d)}_{f} - V^{(d)}_{f=0} = \int\,\frac{d^{d}p}{2\,(2\pi)^{d}}\,\ln\,Q_{f,0}(p) = \qquad  \qquad \nonumber \\
\\
\int\,\frac{d^{d}p}{2\,(2\pi)^{d}}\,\ln\,\left[\frac{I_{-\nu}(p\,\epsilon)}{A_{r_{L}}[I_{-\nu}(pL)]}\cdot \frac{A_{r_{L}}[I_{-\nu}(pL)] + \bar{f}(p)\,A_{r_{L}}[I_{\nu}(pL)]}{I_{-\nu}(p\,\epsilon) + \bar{f}(p)I_{\nu}(p\,\epsilon)}\right], \nonumber
\end{eqnarray}
$\bar{f}(p)$ and $A_{r_{L}}[I_{\pm\nu}(pL)]$ are defined in (\ref{25}) and in (\ref{26}).

For $L=\infty$ (\ref{31}) gives:

\begin{equation}
\label{32}
\tilde{V}^{(d)}_{f,0}(\epsilon,\infty) = \frac{\Omega_{d-1}}{2(2\pi)^{d}\,\epsilon^{d}}\,\int_{0}^{\infty} y^{d-1}dy\,\ln\,\left[\frac{I_{-\nu}(y) (1+\bar{f}(y,\epsilon))}{I_{-\nu}(y) + \bar{f}(y,\epsilon)\,I_{\nu}(y)}\right],
\end{equation}
$y \equiv p\,\epsilon$, $\bar{f}(y,\epsilon) = [2^{2\nu}\Gamma(1+\nu)\,/\, \Gamma(1-\nu)]\,[f\,\epsilon^{2\nu}\,/\,y^{2\nu}]$.

This potential is actually a function of dimensionless double-trace parameter $f\epsilon^{2\nu}$ and may be written as $\tilde{v}_{\epsilon}(f\epsilon^{2\nu})\,/\,\epsilon^{d}$.

For $L < \infty$ potential is a difference of two terms:

\begin{equation}
\label{33}
\tilde{V}^{(d)}(f|\epsilon,L)  = \frac{\tilde{v}_{\epsilon}(f\epsilon^{2\nu})}{\epsilon^{d}} - \frac{\tilde{v}_{L}(fL^{2\nu})}{L^{d}}.
\end{equation}
Thus, this potential can not surve a tool for dynamical fixing of the ratio $\epsilon / L$. This is quite different from the quantum one-loop potential obtained for Robin b.c. at $z = \epsilon$ by Goldberger \& Rothstein - 2000 \cite{Goldb}, and by Carriga \& Pomarol - 2002 \cite{Pomarol}. These authors showed that for integer $\nu$ potential depends on $\ln(L/\epsilon)$ - this gives hope for dynamical explanation of large mass hierarchy.

It may be shown that $\tilde{v}_{\epsilon}(f\epsilon^{2\nu})$ in (\ref{33}) is a monotonic function of its argument. Whereas $\tilde{v}_{L}(fL^{2\nu})$ may have an extremum in some range of values of Robin coefficient $r_{L}$. This question needs further studies.

In any case we may speculate that the double-trace deformation of the boundary CFT $f\,\hat{O}^{2}$ is itself quantum induced from the higher power deformation, say $g\,\hat{O}^{4}$. Then double trace coefficient $f$ may be found in the framework of the AdS/CFT correspondence from a sort of gap equation

\begin{equation}
\label{34}
f = g\,<\hat{O}^{2}>_{f}
\end{equation}
where RHS is found from potential (\ref{33}). This is also a topic for further research.

\subsection{Difference of "regular" and "irregular" potentials. Where is the truth?}

For difference of "regular" ($f = \infty$) and "irregular" ($f = 0$) one-loop quantum potentials it is obtained from (\ref{31}):

\begin{eqnarray}
\label{35}
\tilde{V}^{(d)}_{\infty,0}(\epsilon,L)  = V^{(d)}_{+(L)} - V^{(d)}_{-(L)} =  \frac{\Omega_{d-1}}{2(2\pi)^{d}}\,\int_{0}^{\infty}\,y^{d-1} \cdot \qquad  \qquad \nonumber  \\
\\
\cdot \left\{\,\frac{1}{\epsilon^{d}}\,\ln\left[\frac{I_{-\nu}(y)}{I_{\nu}(y)}\right] - \frac{1}{L^{d}}\,\ln\left[\frac{\left(\frac{d}{2}+r_{L}\right) I_{-\nu}(y)+y\,I_{-\nu}'(y)}{\left(\frac{d}{2}+r_{L}\right)I_{\nu}(y)+y\,I_{\nu}'(y)}\right]\right\}\, dy.   \qquad  \nonumber
\end{eqnarray}

For $L = \infty$ second term is deleted and we get:

\begin{eqnarray}
\label{36}
V^{(d)}_{+} - V^{(d)}_{-} = \frac{\Omega_{d-1}}{2(2\pi)^{d}\epsilon^{d}}\,\int_{0}^{\infty}\,y^{d-1}dy\,\ln\left[\frac{I_{-\nu}(y)}{I_{\nu}(y)}\right] = \qquad \qquad \nonumber \\
 \\
= \frac{2\sin(\pi\nu)\,\Omega_{d-1}}{(2\pi)^{d+1}d\,\epsilon^{d}} \int_{0}^{\infty}\frac{y^{d-1}dy}{I_{\nu}(y)\,I_{-\nu}(y)}.  \qquad  \qquad  \qquad     \nonumber
\end{eqnarray}
Second line results from integration by parts with account that Wronskian $W(I_{\nu}(y),\,I_{-\nu}(y)) = (2\,\sin \pi\nu / \pi)\, y^{-1}$.

Formula (\ref{36}) essentially differs from the expressions for $(V_{+} - V_{-})$ obtained in \cite{Mitra} - \cite{Hartman} - on the one side, and in \cite{Alt2015} and \cite{Saharian} - on the other side. {\bf{Thus we have three answers for one and the same physically meaningful quantity. Where is the truth?}}

\section{\bf{Schwinger-DeWitt expansion}}

\subsection{General expression}

Let us transform bulk scalar field $\Phi(z,x)$ in a way:

\begin{equation}
\label{37}
\Phi (z,x) = (kz)^{(d-1)/2}\, \varphi(z,x),
\end{equation}
and instead of (\ref{2}) we consider differential operator for $\varphi(z,x)$ wherel metric $\tilde{g}_{\mu\nu}(x)$ is introduced and auxiliary "eigenvalue term" $\lambda\,\varphi$ is included:

\begin{equation}
\label{38}
{\bf\hat D}\,\varphi = \left\{\left[-\frac{\partial^{2}}{\partial z^{2}} + \frac{1}{z^{2}}\left(\nu^{2}-\frac{1}{4}\right)\right] + \left[-{\tilde \triangle} + \lambda \right]\right\} \varphi \equiv \{{\bf\hat D}_{z}+{\bf\hat D}_{x}[\tilde{g}_{\mu\nu}]\}\,\varphi.
\end{equation}
Here $\tilde{\triangle} = \tilde{g}^{\mu\nu}(x)\tilde{\nabla}_{\mu}\tilde{\nabla}_{\nu}$. It is evident that ${\rm{Det}}{\bf\hat D}$ (\ref{38}) is equal to ${\rm{Det}}\hat{D}$ (\ref{2}).

${\bf\hat D}_{z}$, ${\bf\hat D}_{x}[\tilde{g}_{\mu\nu}]$ are corresponding differential operators in square brackets in (\ref{38}). Since they commute the heat kernel of differential operator ${\bf\hat D}= {\bf\hat D}_{z}+{\bf\hat D}_{x}$ may be factorized, and conventional Schwinger-DeWitt expansion of the heat kernel of operator ${\bf\hat D}_{x}[\tilde{g}_{\mu\nu}]$ may be put down in a standard way with use of derivatives over auxiliary parameter $\lambda$. Presence of $\lambda$-term in (\ref{38}) means in practice that in all expressions given above in the Report we must change $p \to \sqrt{p^{2} + \lambda}$.

Finally for quantum effective action $\Gamma^{(d)} [\tilde{g}_{\mu\nu}(x)]$ calculated from ratio of determinants of operator ${\bf\hat D}$ (\ref{38}) for two asymptotic b.c. it is obtained:

\begin{eqnarray}
\label{39}
\Gamma^{(d)}_{f_{2}f_{1}} [\tilde{g}_{\mu\nu}(x)] = \int\,d^{(d)}x\,\sqrt{\tilde{g}} \cdot  \qquad  \qquad  \qquad  \nonumber
\\
\\
\cdot \left\{\sum_{n=0}^{\infty}a_{n}(x,x)\left(-\frac{\partial}{\partial\lambda}\right)^{n}\, \int\,\frac{d^{d}p}{2\,(2\pi)^{d}}\,\ln\,Q_{f_{2}f_{1}}(\sqrt{p^{2} + \lambda})\right\}_{\lambda = 0} \nonumber
\end{eqnarray}
where $Q_{f_{2}f_{1}}$ is given in (\ref{28}); $a_{n}(x,x)$ are Schwinger-DeWitt (Gilkey-Seely) coefficients ($a_{0} = 1$, $a_{1} = \tilde{R}^{(d)}/6$, $a_{2} \sim \tilde{R}^{(d)2}$ (symbolically), etc.); and $\lambda$ is an auxiliary "eigenvalue parameter" ($\lambda \equiv \mu^{2}$ of the author's paper \cite{Alt2015} "Sakharov's induced gravity on the $AdS$..."); $\lambda$ must be put to zero after all differentiations over $\lambda$ are fulfilled in (\ref{39}).

\subsection{S-DW expansion: 2-branes RS-model, $d=4$. Induced Planck mass, etc.}

We demonstrate how it works in the case of $d=4$ and for difference of "regular" ($f_{2} = \infty$) and "irregular" ($f_{1}=0$) quantum actions that is for ratio of determinants (\ref{28}):

\begin{equation}
\label{40}
Q_{\infty,0}(\sqrt{p^{2}+\lambda}) = \frac{I_{-\nu}(\sqrt{p^{2}+\lambda}\,\epsilon)}{I_{\nu}(\sqrt{p^{2}+\lambda}\,\epsilon)}\cdot \frac{A_{r_{L}}[I_{\nu}(\sqrt{p^{2}+\lambda}\,L)]}{A_{r_{L}}[I_{-\nu}(\sqrt{p^{2}+\lambda}\,L)]}
\end{equation}
($A_{r_{L}}$ see in (\ref{26})). With this $Q_{\infty,0}$ and for $d=4$ expansion (\ref{39}) takes the form:

\begin{eqnarray}
\label{41}
\Gamma^{(4)}_{\infty,0} [\tilde{g}_{\mu\nu}(x)] = \int\,d^{4}x\,\sqrt{\tilde{g}} \left\{\sum_{n=0}^{\infty}a_{n}(x,x)\left(-\frac{\partial}{\partial\lambda}\right)^{n} \right. \cdot  \frac{\Omega_{3}}{2(2\pi)^{4}\epsilon^{4}} \cdot  \qquad  \qquad  \qquad \nonumber
\\
\\
\cdot \left. \int_{\sqrt{\lambda}\epsilon}^{\infty}\,(y^{2}-\lambda\epsilon^{2})\,y\,dy\,
\ln\left[\frac{I_{-\nu}(y)}{I_{\nu}(y)}\cdot \frac{A_{r_{L}}[I_{\nu}(yL / \epsilon)]}{A_{r_{L}}[I_{-\nu}(yL / \epsilon)]}\right]\right\}_{\lambda = 0} = \qquad \qquad  \nonumber
\\   \nonumber
\\ \nonumber
= \int\,d^{4}x\,\sqrt{\tilde{g}} \left\{\tilde{V}^{(4)}_{\infty,0} + M_{Pl}^{2}\tilde{R} + \frac{1}{4\,g^{2}}\,{\it{O}}(\tilde{R}^{2}) + \sum_{n=3}^{\infty}\,b_{n}\,\frac{1}{[M_{SM}^{2}]^{n-2}}\,{\it{O}}(\tilde{R}^{n})\right\}, \qquad   \nonumber
\end{eqnarray}
where the substitution $y = \sqrt{p^{2} + \lambda} \cdot \epsilon$ was used.

For the first terms ($n = 0, 1, 2$) of $\Gamma^{(4)}_{\infty,0} [\tilde{g}_{\mu\nu}(x)]$ it is obtained:

\begin{equation}
\label{42}
\tilde{V}^{(4)}_{\infty,0} = \frac{\Omega_{3}}{2(2\pi)^{4}}\,\int_{0}^{\infty}\,y^{3}dy\,\left\{\,\frac{1}{\epsilon^{4}}\,\ln\left[\frac{I_{-\nu}(y)}{I_{\nu}(y)}\right] - \frac{1}{L^{4}}\,\ln\left[\frac{A_{r_{L}}[I_{-\nu}(y)]}{A_{r_{L}}[I_{\nu}(y)]}\right]\right\}. \nonumber
\end{equation}

\vspace{0,5cm}

\begin{equation}
\label{43}
M_{Pl}^{2} = \frac{1}{6}\,\frac{\Omega_{3}}{2(2\pi)^{4}}\,\int_{0}^{\infty}y\,dy\, \left\{\frac{1}{\epsilon^{2}}\ln\left[\frac{I_{-\nu}(y)}{I_{\nu}(y)}\right] - \frac{1}{L^{2}}\,\ln\left[\frac{A_{r_{L}}[I_{-\nu}(y)]}{A_{r_{L}}[I_{\nu}(y)]}\right]\right\}. \nonumber
\end{equation}

\vspace{0,5cm}

\begin{eqnarray}
\label{44}
\frac{1}{g^{2}} = \frac{\Omega_{3}}{(2\pi)^{4}}\,\ln\left[\frac{I_{-\nu}(\sqrt{\lambda}\,\epsilon)}{I_{\nu}(\sqrt{\lambda}\,\epsilon)}\cdot  \frac{A_{r_{L}}[I_{\nu}(\sqrt{\lambda}\,L)]}{A_{r_{L}}[I_{-\nu}(\sqrt{\lambda}\,L)]}\right]_{\lambda = 0} =  \nonumber
\\
\\
= \frac{\Omega_{3}}{(2\pi)^{4}}\,\left\{2\nu\,\ln\frac{L}{\epsilon} + \ln\left(\frac{2 + r_{L} + \nu}{2 + r_{L} - \nu}\right) \right\}. \qquad  \qquad \nonumber
\end{eqnarray}

The next one ($a_{3} \sim {\tilde{R}}^{3}$) term in expansion (\ref{41}) looks as:

\begin{equation}
\label{45}
a_{3}(x,x)\,\frac{\Omega_{3}}{4(2\pi)^{4}}\,[a(\nu)\,\epsilon^{2} + b(\nu,r_{L})\,L^{2}],
\end{equation}
where coefficients $a(\nu)$, $b(\nu, r_{L})$ are numbers of order one, they are easily obtained from the standard rows of Bessel functions near zero argument.

Since $L \gg \epsilon$ terms $n \ge 3$ in (\ref{41}) will be of order ${\it{O}}(\tilde{R}^{3})L^{2}$, ${\it{O}}(\tilde{R}^{4})L^{4}$,... ${\it{O}}(\tilde{R}^{n})L^{2(n-2)}$... In RS-model position $L$ of the IR-brane fixes the mass scale of Standard Model $L \cong M_{SM}^{-1}$. Hence Schwinger-DeWitt expansion (\ref{41}) beginning from its $a_{3}$ term is actually an expansion in $[M_{SM}^{2}]^{-1}$. Thus we confirm the result of Section IV of the author's paper \cite{Alt2015} (2015) reached there in essentially more complicated way.

\newpage

\section{Remark in conclusion: integer $\nu$.}

Visible drawback of formulas above is in zero value of potentials for integer $\nu$. This is the difficulty of all the approach based upon different asymptotics at $z \to 0$ of $I_{\nu}$ and $I_{-\nu}$ coinciding at $\nu$ integer.

For $\nu$ integer Green functions of operator $\hat{D}(p)$ (\ref{2}) may be constructed with solutions $z^{d/2}I_{n}(pz)$ and $z^{d/2}K_{n}(pz)$, and there is no problem to build the Green function obeying asymptotic b.c. $\alpha z^{d/2 + n} + \beta z^{d/2 - n}$ where $\alpha = f \beta$:

\begin{equation}
\label{46}
G(z,z') \sim \frac{[K_{n}(pz) + \bar{f}_{n}(p) I_{n}(pz)]\,K_{n}(pz')\,\theta(z'-z)+ (z \leftrightarrow z')\}}{\bar{f}_{n}(p)}
\end{equation}
(we consider here $L = \infty$ and exclude special case $n=0$),

\begin{equation}
\label{47}
\bar{f}_{n}(p) = 2^{2n-1}n!\,(n-1)!\,\frac{f}{p^{2n}}
\end{equation}
in analogy with $\bar{f}(p)$ of Gubser \& Mitra \cite{Mitra} - see (\ref{6}).

And there is no problem to obtain with Barvinsky-Nesterov method simple expressions for quantum determinants - quantum potentials.

"Regular" asymptotic b.c. corresponds here to $f=\infty$. {\bf{But what value of the double-trace coefficient $f$ corresponds in $K_{n}(pz) + \bar{f}_{n}I_{n}(pz)$ to "irregular" b.c.? The answer is unclear for me.}}

\section*{Acknowledgements} Author is grateful to Andrei Barvinsky, Ruslan Metsaev, Dmitry Nesterov and Arkady Tseitlin for fruitful discussions and helpful critical notes.

\end{document}